


\input phyzzx

\def\H{\hat{H}}

\def\SUMrs{\Sigma_{\bf r,\sigma}}
\def\SUMrrp{\Sigma_{({\bf r,r^\prime})}}
\def\SUMrrps{\Sigma_{({\bf r,r^\prime}),\sigma}}

\def\SUMk{\Sigma_{\bf k}}
\def\SUMks{\Sigma_{\bf k,\sigma}}

\def\eipr{e^{i\bf \vec{\pi} r}}

\def\Crs{\hat{c}_{\bf r,\sigma}}

\def\Cru{\hat{c}_{\bf r,\uparrow}}
\def\Crd{\hat{c}_{\bf r,\downarrow}}
\def\Crps{\hat{c}_{\bf r^\prime,\sigma}}
\def\Crpu{\hat{c}_{\bf r^\prime,\uparrow}}
\def\Crpd{\hat{c}_{\bf r^\prime,\downarrow}}

\def\Crsd{\Crs^\dagger}

\def\Crud{\Cru^\dagger}
\def\Crdd{\Crd^\dagger}

\def\D{\Delta}
\def\Drrp{{\Delta}_{\bf r,r^\prime}}
\def\Drrpd{\Drrp^\ast}
\def\hDrrp{\hat{\Delta}_{\bf r,r^\prime}}
\def\hDrrpd{\hDrrp^\dagger}

\def\K{\chi}
\def\Krrp{{\chi}_{\bf r,r^\prime}}
\def\Krrpd{\Krrp^\ast}
\def\hKrrp{\hat{\chi}_{\bf r,r^\prime}}
\def\hKrrpd{\hKrrp^\dagger}

\def\Cks{\hat{c}_{\bf k,\sigma}}

\def\Cku{\hat{c}_{\bf k,\uparrow}}
\def\Ckd{\hat{c}_{\bf k,\downarrow}}

\def\Cpku{\hat{c}_{\bf \vec{\pi}-k,\uparrow}}

\def\Cksd{\Cks^\dagger}

\def\Ckud{\Cku^\dagger}
\def\Ckdd{\Ckd^\dagger}

\def\Cpkud{\Cpku^\dagger}

\def\Gks{\hat{\gamma}_{\bf k,\sigma}}
\def\Gku{\hat{\gamma}_{\bf k,\uparrow}}
\def\Gkd{\hat{\gamma}_{\bf k,\downarrow}}

\def\Gpku{\hat{\gamma}_{\bf \vec{\pi}-k,\uparrow}}

\def\Gksd{\Gks^\dagger}
\def\Gkud{\Gku^\dagger}
\def\Gkdd{\Gkd^\dagger}

\def\Gpkud{\Gpku^\dagger}

\def\X{X_{\bf k}}
\def\Y{Y_{\bf k}}
\def\F{F_{\bf k}}
\def\G{G_{\bf k}}

\overfullrule=0pt
\nopubblock

\titlepage
\hoffset=0in
\hsize=6.5in
\voffset=0in
\vsize=8.9in
\rightline{UCDPHYS-PUB-20-95}
\rightline{June, 1995 \hskip .85in}

\doublespace

\title{ ORVB Mean-Field Calculation in the Tight-Binding Model
with Anti-Ferromagnetic Exchange}

\author{Ling-Lie Chau$^\ast$ and Ding-Wei Huang$\dagger$}

\address{\it $^\ast$ Department of Physics, University of California,
Davis, CA~~95616;\break
$^\dagger$ Department of Physics, Chung Yuan Christian University, Chung-Li,
Taiwan.}

\bigskip
\bigskip

\abstract

We give a mean-field calculation for the odd-resonating-valence-bond ORVB
pairing scheme.
We obtain interesting  quasi-particle excitation energy $E_{\bf k}$ as a
function of momentum
${\bf k}$. It is distinctively different from those of the
$d_{x^2-y^2}$-wave,
 the anisotropic-s-wave, and the p-wave. It is a gapless theory for
superconductivity with well defined
Fermi surface. The ground state of the  ORVB scheme
is not an eigenstate of the parity or the time-reversal transformation, thus
both symmetries
are violated. Some of them are already manifested in $E_{\bf k}\neq E_{-{\bf
k}} $. It is interesting to find out
if such pairing order-parameter scheme exits in some materials  in nature.

\vfill
\endpage

The discovery of high $T_c$ superconductivity\attach{[1]} led to an enormous
interest
in models with strong correlations, especially the
Hubbard model\attach{[2]} with strong repulsion.\attach{[3-5]}
In a previous paper\attach{[6]}, we pointed out  the existence of extended
local symmetries in the large-$U$ Hubbard model at half-filling
as well as the anti-ferromagnetic exchange model,
and proposed a new order-parameter scheme, the
odd-resonating-valence-bond (ORVB)
 scheme.
We later in Ref.[7] showed that, besides the current popular
$d_{x^2-y^2}$-wave
order-parameter scheme, the ORVB scheme is another natural consistent
solution
to the rigorous constraints imposed on the system under the thermal
equilibrium.
(Earlier Zhang pointed out that the pure s-wave order-parameter scheme is not
allowed.\attach{[8]})

In this paper, we report the result of a mean-field calculation for the ORVB
pairing scheme.
We obtain interesting  quasi-particle excitation energy $E_{\bf k}$
as a function of the momentum
${\bf k}$. It is a gapless theory for superconductivity with well defined
Fermi surface.
They are distinctively different from those of the $d_{x^2-y^2}$-wave
pairing,
 the anisotropic-s-wave pairing, the p-wave pairing, and many other pairing
schemes.\attach{[9]}
Our ORVB scheme can be tested in the
angle-resolved-photo-emission-spectroscopy (ARPES) experiments,
however the experiments must scan many key directions in the whole Brillouin
zone.
The ground state of the  ORVB scheme
is not an eigenstate of the parity or the time-reversal operator, thus both
symmetries
are violated. Some of these symmetry violation effects have
already manifested in $E_{\bf k}\neq E_{-{\bf k}} $.

It is interesting to find out experimentally
if  such pairing order-parameter scheme exits in some materials  in nature or
if it is a scheme for some high $T_c$ superconductivity .

We start with the following mean-field Hamiltonian of the tight-binding
model with anti-ferromagnetic exchange,
$$\eqalignno{
\H_{mean}^{ORVB}=&t\SUMrrps(\Crsd\Crps+h.c.)+\mu\SUMrs\Crsd\Crs   \cr
-&(J/2)\SUMrrp\{\Drrp\hDrrpd+\Drrpd\hDrrp
+\Krrp\hKrrpd+\Krrpd\hKrrp\},&(1)
}$$
where
$$
\hDrrp\equiv \Cru\Crpd-\Crd\Crpu ~,~~~ \hKrrp\equiv \Crud\Crpu+\Crdd\Crpd ~.
\eqno(2)
$$
These two operators are related by the $SU(2)$ symmetry, which
is the symmetry of the $J$-term or the symmetry of the anti-ferromagnetic
exchange.\attach{[6]}
In obtaining the  mean-field Hamiltonian, we have used the ORVB order
parameters:
 $\Drrp\equiv <\hDrrp> \equiv Tr( e^{\beta\H} \hDrrp)$,
where $\H$ is the Hamiltonian and $\beta =(kT)^{-1}$.
At temperature $T=0$, $<\hDrrp>=<0| \hDrrp |0>$, i.e., the expectation value
of the ground state; similarly for
$\Krrp\equiv <\hKrrp>$. The
doping parameter $\delta$ is defined by
$
1-\delta  =  N^{-2} ~\SUMrs <\Crsd\Crs>.
$

The ORVB order-parameters can be parameterized in the following way:
$$\eqalignno{
\Drrp=&\delta_{\bf r',r\pm e} ~\eipr~ \D e^{ i\theta} ,&(3)\cr
\Krrp=&\delta_{\bf r',r \pm e} ~\K e^{\pm i\phi} ,&(4)
}$$
where the unit vector ${\bf e=e_x}$ or ${\bf e_y}$. As shown in Ref.[7],
the rigorous constrains imposed by thermal equilibrium
allow all these real order parameters,
$\D, \theta, \K$ and $\phi$  to be nonzero and unequal in different
directions. For simplicity, as done in many other order parameter schemes,
we  choose them to be constants, independent of sites ${\bf r}$.
Notice that from the definitions of $\hDrrp$ and $\hKrrp$ in Eq.(2), we have
$
\D_{\bf -r,-r'}=-\Drrp~$
and
 $~\K_{\bf -r,-r'}=(\Krrp)^\ast ~.
$
In the case $\D =0$ but $\K \neq 0$ and $\phi \neq 0$, the ORVB
scheme becomes the flux phase scheme.\attach{[10]} However the main new
features of
the ORVB scheme come from  $\D \neq 0$ and the interplay between the two sets
of order parameters given in Eqs. (3) and (4).

After the Fourier transformation, the Hamiltonian in momentum space becomes
$$\eqalignno{
\H_{mean}^{ORVB}=&\SUMks\{2t C_k+\mu
-J\K [cos(k_x+\phi)+cos(k_y+\phi)]\}\Cksd\Cks \cr
&-\SUMk\{J\D S_k[ e^{-i(\theta+{\pi\over 2})} \Cpku\Ckd
+ e^{i(\theta+{\pi\over 2})} \Ckdd\Cpkud] \},&(5)
}$$
where $C_k \equiv cosk_x+cosk_y$ and $S_k \equiv sink_x+sink_y$. Obviously
from
 Eq. (5) we see that the momenta of the electrons in the pairing are
 ${\bf k}$ and ${\bf\vec{\pi} -k}$,
in contrast to the ${\bf k}$ and ${\bf -k}$ of the BCS pairing.\attach{[11]}

The Hamiltonian $\H_{mean}^{ORVB}$ can be diagonalized by the following
Bogoliubov transformation
$\Ckd= ~u_k^*~ \Gkd +v_k~ \Gpkud ~$
 and
$\Cpkud=-v_k^*~ \Gkd +u_k~ \Gpkud ~,$
 where $\Gks$  and $\Gksd$ are quasi-particle operators.
Then the Hamiltonian becomes
$$
\H_{mean}^{ORVB}=\SUMks (E_{\bf k} ^{ORVB}\Gksd\Gks +E_{\bf k}^0) ~~,
\eqno(6)
$$
with the energy given by
$$\eqalignno{
E_{\bf k}^{ORVB}=&(2t-J\K cos\phi) C_k
\pm\sqrt{[J\D S_k]^2+[\mu +J\K S_k sin\phi]^2}~~,
  &(7)
}$$
where $\pm$ are taken as the sign of $[\mu+J\K S_k sin\phi]$.
 Notice that when the order parameters become zero,
$
E_{\bf k}^{ORVB}=\epsilon_{\bf k} \equiv 2t C_k +\mu~.
$

The ORVB energy expression, Eq. (7),  is quite different from those of the
BCS-type
of theories,
$
E_{\bf k}^{BCS}=\pm\sqrt{ {\epsilon_{\bf k}}^2+[J\D_{\bf k}]^2},
$
in which $\D_{\bf k} =\D_s$ for the s-wave scheme  (the original BCS
order parameter);
$\D_{\bf k} =\D_d (cosk_x -cosk_y)$ for the $d_{x^2-y^2}$-wave scheme;
$\D_{\bf k} =\D_a [(cosk_x -cosk_y)^4 +$ {\it constant}]
for the anisotropic s-wave scheme; (the sign in front of $
E_{\bf k}^{BCS}$ is that of $\epsilon_{\bf k}$).
 Notice that when the order parameters are zero (at  $T > T_c$),
the ORVB scheme has the same energy expression, $
E_{\bf k}=\epsilon_{\bf k}$, as the BCS-type of theories.
When the order parameters becomes nonzero (at  $T < T_c$),
the BCS-type of theories gives energy discontinuity (gap) around
$\epsilon_{\bf k}=0$.
 The ORVB scheme gives very different energy discontinuities, which we shall
discuss later.

Notice that the momentum-factor associated with $\D$ in the $E_{\bf
k}^{ORVB}$ of
Eq.(7)
is $S_k$, a result of the pairing momentum being $\pi$.
The momentum-factor associated with $\K$ is $cos(k_x+\phi) +cos(k_y+\phi)$.
It decomposes into
$ C_k  cos\phi $ and $S_k sin\phi$;
the former
gives the term outside the square root in Eq.(7),
and the latter gives the term inside the square root in Eq.(7).

The ground state is defined by
$\Gks |0> =0 ~$ for $~ E_{\bf k} > 0 ~$ and
$\Gksd |0> =0 ~$ for $~ E_{\bf k} < 0. $
The quasi-particle operators $\Gksd$ and $\Gks$ are free fermions.
$<\Gksd\Gks>$ are the only non-vanishing expectation-values.
The requirement that the occupation number at a temperature $T$
is given by the Fermi-Dirac distributions
$
< \Gksd\Gks > =~
[ e^{\beta E_{\bf k}} +1 ]^{-1}~
$
 imposes a set of consistent conditions upon the pairing
parameters.
After Fourier-transforming to momentum space and
then Bogoliubov-transforming $\hat{c}$ and $\hat{c}^{\dagger}$ to
$\hat{\gamma}$ and $\hat{\gamma}^{\dagger}$,
we obtain the following four equations:
$$
\delta =N^{-2}~\SUMk~
[J\K S_k sin\phi~+\mu]\F \X^{-1},
\eqno({\bf 8})
$$
$$\eqalignno{
2 J\K^2 cos^2\phi -4 t\K cos\phi &=
 N^{-2}~\SUMk~(\Y~\G),
&({\bf 9})\cr
2 J~(\D^2+\K^2 sin^2\phi) +\mu\delta &=
 N^{-2} ~\SUMk~(\X~\F),
&({\bf 10})\cr
2J(\D^2 -\K^2sin^2\phi) -\mu\delta &=\cr
N^{-2} ~\SUMk~ \{[J\D S_k]^2-[&J\K S_k sin\phi+\mu]^2\}\F \X^{-1},
&({\bf 11})\cr}
$$
where
$$\eqalignno{
\X& \equiv \sqrt{[J\D S_k]^2 +[J\K S_k sin\phi +\mu]^2},&(12a)\cr
\Y& \equiv (J\K cos\phi-2t\delta) C_k,&(12b) \cr
\F& \equiv sinh(\beta \X) [ cosh(\beta \X)+cosh(\beta \Y)]^{-1},&(12c)\cr
\G& \equiv sinh(\beta \Y) [ cosh(\beta \X)+cosh(\beta \Y)]^{-1}.&(12d)}$$

For given temperature $\beta$ and doping parameter $\delta$, totally there
are four
equations, Eqs.(8), (9), (10), and (11), for the four unknowns $\D$, $\K$,
$\phi$, and $\mu$.
(Notice that the phase $\theta$ does not appear in these equations.)

At $T=0$, the temperature dependent factor
$\F=1$ for $\X>|\Y|$ and $\F=0$ for $\X<|\Y|$;
$\G=1$  for $\Y>\X$, $\G=-1$  for $\Y<-\X$, and $\G=0$  for $|\Y|<\X$.
Therefore
Eqs.(8), (9), (10), and (11) are further simplified, which can be trivially
obtained
and we do not write them out explicitly.

We have analyzed Eqs.(8), (9), (10), and (11) numerically at small values of
the doping parameter
$\delta$
at various values of $T$.
 For a given small value of
$\delta$, we find that the solution for these equations giving a small value
of $\mu$ and
Eq.(11) does not impose a strong independent condition from that of Eq.(10).
[These can be understood by the fact that at $\delta=0$ and $\mu=0$
 Eq.(11) reduces to Eq.(10) and Eq.(8) becomes an identity.  However, our
solution is different from the one exactly
at half-filling, i.e. $\delta=0$ and $\mu=0$, where further degeneracy
happens and gives a different
solution.]
Therefore for a small doping parameter $\delta$ only two of the parameters,
$\K^2 cos^2\phi$ and $\D^2+\K^2 sin^2\phi$
 are determined, essentially by Eq.(9) and Eq.(10).
The approximate solutions we have obtained are the following:
the critical temperature $T_c$ (that we define to be the temperature below
which all
the order parameters are nonzero) and the order parameters at $T=0$
are all of the order of the interaction strength $J$;
the order-parameters decrease as the doping parameter $\delta$ and/or as the
temperature $T$
increases;
$T_c$ decreases with the increasing value of the doping parameter $\delta$.
For example, at $T=0$ we find the approximate solution
 at $\delta\simeq 0.18$ ($t\simeq 0.18$):
 $\mu\simeq 0.08$,  $  \D^2+\K^2 sin^2\phi\simeq 0.032$, and   $\K^2
cos^2\phi \simeq 0$.
{}From these discussions, we also see that our ORVB scheme contains
the flux phase (i.e. $\K \ne 0$  and $\D = 0$) and the case of $\K = 0$  and
$\D \ne 0$ as special cases. We shall see below
that it is the general cases of both $\K \ne 0$  and $\D \ne 0$ that give the
interesting new physical phenomena.

With these solutions, we can calculate the quasi-particle excitation energy
$E_{\bf k}^{ORVB}$
 of Eqs.(6) and (7).
In Fig.1 we show the $E_{\bf k}^{ORVB}$ distribution as function of ${\bf k}$
 for
$\D\simeq 0.17$ and $\K sin\phi\simeq -0.06.$
Notice that the lattice lines of the Brillouin zones are no more the
symmetry lines.
We shall show later, this is a result from the violation of the parity
as well as the time reversal symmetries of our ORVB pairing scheme.
The $\K sin\phi\ne 0$ gives a wall of discontinuity (gap) in the ${\bf
k}$-plane
specified by
$
S_k=-\mu  /{(J\K sin\phi)},
$
surrounding the centers
at $(k_x,k_y)=(\pi/2,\pi/2)$, modulus $2\pi$ in both $k_x$ and $k_y$.
The magnitude of the discontinuity is $|(2\mu\D)/(\K sin\phi)|$.
Notice in Fig. 1 there are two lines of zero energy (gaplessness).
\attach{[12]}The line of
 $E_{\bf k}=\epsilon_{\bf k}$ is also symmetrically centered around
$(k_x,k_y)=(\pi/2,\pi/2)$, which is not shown in the figure but can be easily
seen from Eq.(7).

The quasi-particle excitation energy
$E_{\bf k}$ distributions for the BCS-type theories are very different from
that of the ORVB scheme we just presented. For BCS-type theories, the lattice
lines of the Brillouin zones are the symmetry lines, a result of parity and
time reversal symmetries. The anisotropic s-wave scheme is a finite-gap
theory (i.e. $E_{\bf k} \ne 0$ in the whole Brillouin zone) and the
$d_{x^2-y^2}$-wave scheme has isolated four points of gaplessness. We shall
mention at the end of the paper how these differences can be experimentally
distinguished.

At $T=0$, the occupation number in the quasi-particle space is
$n_{\bf k}^\gamma \equiv$  $<{\Gkud\Gku +\Gkdd\Gkd}>$ $ =0~$ for $~ E_{\bf k}
> 0 ~$ and
$n_{\bf k}^\gamma =2~$ for $~ E_{\bf k} < 0 ~$.
After Bogoliubov-transforming $\hat{\gamma}$ and $\hat{\gamma}^{\dagger}$
back to $\hat{c}$ and $\hat{c}^{\dagger}$,
we obtain the occupation number in the electrons
$n_{\bf k}^e \equiv <{\Ckud\Cku +\Ckdd\Ckd}>$, shown in Fig.2. These
occupation number distributions are also very different from those of the
BCS-type of theories.

We note that for every $\K sin\phi < 0$ solution,
$\K sin\phi =-0.06$ in Fig.1,
there is another equally allowed ORVB solution with $\K sin\phi >0$
with the discontinuity
center changed to $(k_x,k_y)=(-\pi/2,-\pi/2)$.
The other characteristic are the same as those just discussed and presented
for
$\K sin\phi <0$.

Besides the ORVB,
there is another kind of solution
to the thermal equilibrium constraint equations,\attach{[7]}
which we call the (ORVB)' pairing scheme.
In stead of Eqs.(3) and (4), we have
$$\eqalignno{
\Drrp=&\delta_{{\bf r',r+e_x}} ~\eipr\D ~e^{i\theta}
     =-\delta_{{\bf r',r+e_y}} ~\eipr\D ~e^{i\theta} ~~,&(13)\cr
\Krrp=&\delta_{{\bf r',r+e_x}} ~\K ~e^{i\phi}
     =-\delta_{{\bf r',r+e_y}} ~\K ~e^{i\phi} ~~,&(14)
}$$
and $\D, \K, \theta$, and $\phi$ are real constants.
The net result is that
for $\K sin\phi <0$,
the discontinuity shifted to surround the centers at
$(k_x,k_y)=(\pi/2,-\pi/2)$ modulus $2\pi$ in both $k_x$ and $k_y$;
for $\K sin\phi >0$, the discontinuity shifted to surround the centers
at $(k_x,k_y)=(-\pi/2,\pi/2)$ modulus $2\pi$ in both $k_x$ and $k_y$.

Next, we discuss the intrinsic parity and time reversal symmetries.
Both the $\Drrp$ and $\Krrp$, Eqs.(3) and (4),
violate the parity and the time reversal symmetries.

Under parity transformation $\hat{P}$,
for any eigenstates of parity $|a>$ and $|b>$,
the expectation value of any operator $\hat{O}$
has the identity
$<a|\hat{O}|b> = <a|\hat{P}\hat{O}\hat{P}^{-1}|b> ~.$
{}From the fact
$\hat{P}\hat{c}_{\bf r,\sigma}\hat{P}^{-1}=\hat{c}_{-\bf r,\sigma}$
and
$\hat{P}\hat{c}_{\bf r,\sigma}^\dagger\hat{P}^{-1}
=\hat{c}_{-\bf r,\sigma}^\dagger$, we obtain
$\hat{P}\hDrrp\hat{P}^{-1}=\hat{\D}_{{\bf -r,-r'}}$ and
$\hat{P}\hKrrp\hat{P}^{-1}=\hat{\K}_{{\bf -r,-r'}}$ from the definitions of
these
operators in Eq.(4).
If the eigenstates of Hamiltonian are eigenstates of the parity operator, the
first
relation implies
$< \hDrrp > = < \hat{\D}_{{\bf -r , -r'}} >$; in which the right-hand-side
(r.h.s.)
 is $\Drrp$,  and the left-hand-side (l.h.s.) is $-\Drrp$, using Eqs.(3) and
(4);
 thus must $\Drrp =0$. The second relation implies
$< \hKrrp > = < \hat{\K}_{{\bf -r , -r'}} >$, in which the r.h.s. is  $\Krrp$
and the
l.h.s. is $\Krrp^\ast$; thus must $\Krrp =$ real, i.e., $\phi =0$.
Therefore, $\D \ne 0$ or the complexity of $\K e^{i\phi}$, i.e., $\phi\ne 0$,
implies that eigenstates of the Hamiltonian can not be the eigenstates of
parity, i.e.,
parity is violated. Notice that this parity violation property is different
from that of the
p-wave pairing scheme, in which the ground state is an odd eigenstate of the
parity
transformation.\attach{[9]}

Under time reversal transformation $\hat{T}$,
$\hat{T}|a>=|a_t>$ and $\hat{T}|b>=|b_t>$,
the expectation value of any operator $\hat{O}$
has the identity
$<a|\hat{O}|b> = <a_t|\hat{T}\hat{O}\hat{T}^{-1}|b_t>^\ast ~.$
 From the fact that
$\hat{T}\hat{c}_{\bf r,\sigma}\hat{T}^{-1}=\hat{c}_{\bf r,-\sigma}$
and
$\hat{T}\hat{c}_{\bf r,\sigma}^\dagger\hat{T}^{-1}
=\hat{c}_{\bf r,-\sigma}^\dagger$,
we obtain $\hat{T}\hDrrp \hat{T}^{-1} = -\hDrrp $
and $ \hat{T} \hKrrp \hat{T}^{-1} =\hKrrp$ from the definitions of these
operators in Eq.(4).
If the ground state is the eigenstate of time reversal $\hat{T}|0>=|0>$,
$<0| \hDrrp |0> = <0| -\hDrrp |0>^\ast $ and
$<0| \hKrrp |0> = <0| \hKrrp |0>^\ast$; which implies at temperature $T=0,$
$\Drrp =-\Drrp^\ast$ (i.e., $\Drrp$ is pure imaginary and $\theta =\pi/2$)
and
$\Krrp =\Krrp^\ast$ (i.e., $\Krrp$  is real and $\phi=0, \pi$).
Thus $\D cos\theta\ne 0$ or $\K sin\phi\ne 0$  implies the violation
of time reversal symmetry.
(Note that $\K e^{i\phi}=\K$, being real, is not an interesting order
parameter,
Since it only renormalizes the tight-binding term in the Hamiltonian.)

If the ground state is an eigenstate of time reversal $\hat{T}|0>=|0>$,
following similar discussions as above, we can easily show that
 $E_{\bf k}\equiv <0| \H_{mean} |0>=E_{\bf -k}$
from either the Hermiticity of the Hamiltonian,
$\H_{mean}^\dagger=\H_{mean}$,
or $\hat{T}\H_{mean}\hat{T}^\dagger=\H_{mean}$.
Therefore, $E_{\bf k} \ne E_{\bf -k}$ indicates that the ground state is not
an
eigenstate of time reversal.
In our case, $E_{\bf k}\ne E_{\bf -k}$ comes solely from the $\K sin\phi$
term.
If $\K sin\phi =0$, our $E_{\bf k}=E_{\bf -k}$ even if there is
 time-reversal violation from  $\D cos\theta\ne 0$; thus
the time reversal violation effect from $\D cos\theta$
can not be observed in $E_{\bf k}$, since $E_{\bf k}$ is not
sensitive to the phase of $\D e^{i\theta}$.

Finally we point out some distinct features of the ORVB scheme that
experiments can look for. In our ORVB scheme the wall of the energy
discontinuity
is off-centered
in the Brillouin zone, centered in one of the quadrant of the Brillouin zone
(see Fig. 1 and
discussions in the text around the Eqs.(12) to (14)), in contrast to those of
the $d_{x^2-y^2}$-wave and the anisotropic
 s-wave which are all centered symmetrically around the center of the
Brillouin zone.
It will be interesting to study such asymmetric distributions
in experiments that can measure the
 energy discontinuities, like the ARPES experiments. \attach {[13]}  Due to
complexity of
the order parameters, in experiments that can measure the phases of order
parameters
(like the Josephson interference experiments) the ORVB scheme   can give
interference
phase shifts different from $\pi$, which is the only kind of interference
phase shift
 the $d_{x^2-y^2}$-wave scheme can produce (due to the reality of its order
parameter).
 We have also calculated the temperature dependence of the penetration depth.
It increases with temperature  faster than that from the $d_{x^2-y^2}$-wave
case.
(We shall publish our detail numerical results in a separate paper.)

To conclude, our ORVB scheme offers a pairing order parameter
scheme for superconductivity very different from the current popular models,
e.g.,
 the $d_{x^2-y^2}$-wave scheme, the anisotropic s-wave scheme, and the p-wave
scheme.
 It has interesting and distinct features that can be experimentally tested.
Seeing the  ORVB  characteristics in some
materials will certainly be very exciting.

\bigskip
\bigskip

\noindent{\bf ACKNOWLEDGMENT}

We would like to thank T. Devereaux and R. R. P. Singh
for sharing with us their expert knowledge in the field and for helpful
discussions; and Z.-X. Shen for informative discussions on the ARPES
experiments.
L.-L. Chau would like to thank P. W. Anderson and D. Pines
for generously making time available (on separate occasions long time ago
before
their recent public debate)  to share with her their views on high $T_c$
superconductivity.
D.-W. Huang acknowledges the hospitality from the
Physics Department of the University of California at Davis,
where the work was done.

This research is supported in part by the US Department of Energy.

\bigskip

\noindent{\bf FIGURE CAPTIONS}

\noindent
{\bf Fig.1}~
The equal  $E_{\bf k}$, the quasi-particle energy, contour diagram
as function of the momentum ${\bf k}$ in the ORVB scheme
with a small doping parameter $\delta$. For the values of parameters used and
discussions,
 see the text. Notice that the wall of discontinuity (gap) is centered around
the center of a quadrant of
the Brillouin zone. Shown  in the figure is the case of the center being
at $(k_x,k_y)=(\pi/2,\pi/2)$.
It is equally likely in the ORVB scheme that the center is at
$(-\pi/2,\pi/2)$,
$(\pi/2,-\pi/2)$, or $(-\pi/2,-\pi/2)$.

\noindent
{\bf Fig.2}~
The three dimensional view of the electron density $n^e_{\bf k}$ of the ORVB
scheme.

\vfil
\endpage

\noindent{\bf REFERENCES}

\noindent
{\bf [1]~}
J. G. Bednorz and K. A. Muller, Z. phys. \underbar{B64} (1986) 189;
M. K. Wu, J. R. Ashburn, Y. Q. Wang, and C. W. Chu, Phys. Rev. Lett.
\underbar{58} (1987) 908.

\noindent
{\bf [2]~}
J. Hubbard, Proc. Roy. Soc. \underbar{A276} (1963) 238;
P. W. Anderson, Science \underbar{235} (1987) 1196.

\noindent
{\bf [3]~}
G. Baskaran, Z. Zou, and P. W. Anderson,
Solid State Commun. \underbar{63} (1987) 973;
V. Emery, Phys. Rev. Lett. \underbar{58} (1987) 2794;
G. Kotliar, Phys. Rev. \underbar{B37} (1988) 3664.

\noindent
{\bf [4]~}
S. Chakravarty, A. Sudb{\o}, P. W. Anderson, and S. Strong,
Science \underbar{261} (1993) 337.

\noindent
{\bf [5]~}
P. Monthoux and D. Pines,
Phys. Rev.  \underbar{B47} (1993) 6069;
D. Pines,
Physica  \underbar{B199 and 200} (1994) 300; D. J. Scalapino, Physics Reports
\underbar{250}(1995) 329; J. R. Schrieffer, Sol. State Comm. \underbar{92}
(1994) 129.

\noindent
{\bf [6]~}
L.-L. Chau, D.-W. Huang, and Y. Yu,
Phys. Rev. Lett. \underbar{68} (1992) 2539.

\noindent
{\bf [7]~}
L.-L. Chau and D.-W. Huang,
{\it Constraints on Order Parameters and Correlation Functions
of Systems in Thermal Equilibrium}, submitted to Phys. Rev. B.

\noindent
{\bf [8]~}
S. C. Zhang, Phys. Rev. \underbar{B42} (1990) 1012.

\noindent
{\bf [9]~}
For example the OPSP pairing scheme of R. T. Scalettar, R. R. P. Singh, and
S. C. Zhang,
 Phys. Rev. Lett. \underbar{67}  (1991) 370; the p-wave pairing scheme of
A. V. Balatsky, E. Abraham, D. J. Scalapino, and J. R. Schrieffer, Physica B
(1994) 363.

\noindent
{\bf [10]~}
I. Affleck and J. B. Marston,
Phys. Rev. \underbar{B37} (1988) 3774.

\noindent
{\bf [11]~}
J. Bardeen, L.N. Cooper, and J.R. Schrieffer,
Phys. Rev. \underbar{108} (1957) 1175.

\noindent
{\bf [12]~}
As $\K sin\phi \rightarrow 0$, the discontinuity disappears and
the energy $E_{\bf k}$ becomes continuous:
$
E^{ORVB}_{\bf k}=(2t-J\K cos\phi) C_k
 +\sqrt{[J\D S_k]^2 +\mu^2}.
$
As $|\K sin\phi |$ increases (remember that
$\D^2 +\K^2 sin^2\phi \simeq$ constant), the area
enclosed by the discontinuity
enlarges and the discontinuity along the boundary becomes smaller.
When $|\K sin\phi |$ reaches its maximum ($\D =0$), the spectrum becomes
continuous
again,
$
E_{\bf k}^{ORVB}=(2t-J\K cos\phi) C_k
+\mu +J\K S_k sin\phi.
$

\noindent
{\bf [13]~}
It is impressive that the energy discontinuities can be measured at
all.
So far only partial regions of the Brillouin zone of Bi2212 has been scanned
for the energy discontinuities by the ARPES experiments, see
Z.-X. Shen, {\it et al.}, Science \underbar{267} (1995) 343, and references
therein.

\vfill
\eightrm
\singlespace
qc;manuscripts;orvb;
0\_orvb\_tb\_cond.tex
llc :6/30/95
\endpage

\end